\documentclass[10pt,aps,twocolumn,pre,floatfix,superscriptaddress]{revtex4-1}
\usepackage{graphicx}
\usepackage{epstopdf}
\usepackage{amssymb,amsfonts,amsmath}
\usepackage{color}
\usepackage{dcolumn}
\usepackage{bm}
\usepackage[caption=false]{subfig}
\usepackage{wrapfig} 
\usepackage{xcolor}
\usepackage{mdframed}
\usepackage[many]{tcolorbox}
\usepackage{lmodern}
\usepackage{lipsum}
\usepackage{amsmath, nccmath}
\usepackage{multirow}
\usepackage{tabu}
\usepackage{float}
\usepackage{gensymb}
\usepackage{capt-of}
\begin{document}

\title{ Zeros of partition function for Continuous Phase Transitions
  using cumulants}

\author{Debjyoti Majumdar}
\email{debjyoti@iopb.res.in}
\affiliation{Institute of Physics, Bhubaneswar, Orissa, 751005, India}
\affiliation{Homi Bhabha National Institute, Training School Complex, Anushakti Nagar, Mumbai 400085, India}
\author{Somendra M. Bhattacharjee}
\email{somen@iopb.res.in}
\affiliation{Institute of Physics, Bhubaneswar, Orissa, 751005, India}

\date{\today}

\begin{abstract}
  This paper explores the use of a cumulant method to determine the zeros of partition functions for continuous phase transitions.  Unlike a first-order transition, with a uniform density of zeros  near the transition point, a continuous transition is expected to  show a power law dependence of the density with a nontrivial slope  for the line of zeros.  Different types of models and methods of  generating cumulants are used as a testing ground for the method.  These include exactly solvable DNA melting problem on hierarchical  lattices, heterogeneous DNA melting with randomness in sequence,  Monte Carlo simulations for the well-known square lattice Ising  model.  The method is applicable for closest zeros near the  imaginary axis, as these are needed for dynamical quantum phase  transitions. In all cases, the method is found to provide the basic
  information about the transition, and most importantly, avoids root  finding methods.
\end{abstract}

\maketitle
\section{\label{sec:level1}Introduction}
A phase transition is generically defined as a point of non-analyticity of any quantity of interest, like the free energy for an equilibrium system, and developing an understanding of phase transitions is associated with keywords like critical exponents, order parameter, correlation length, etc., and, not the least, the transition point \cite{goldenfeld}.  One of the innovative ways of locating and finding the nature of a transition is the method of finding the zeros of the partition function by allowing the
temperature or similar intensive field like quantities to be complex. This method was first incorporated by Lee and Yang.  The failure of the cluster expansion method by Mayer and coworkers in describing the properties of the liquid phase, motivated Lee and Yang to study the liquid-gas phase transition in the complex fugacity plane and the two-dimensional (2d) Ising model in the complex magnetic field plane \cite{leeyang01,leeyang02}.  Of late, studies of zeros of partition functions, or its equivalents, have become relevant in many fields, like quantum dynamics, polymers, QCD, and analysis of experimental data, to name a few \cite{heyl,rensburg,nagata}.  The method of Lee-Yang zeros are not only confined to cases of equilibrium systems but have also been applied to systems out of equilibrium successfully \cite{evans}.  Although it is difficult to realize complex parameters in experiments, recently there have been attempts to explore the Lee-Yang zeros of an
Ising type spin bath \cite{wei, peng,brandner}.

The basic idea of the method of zeros is to find the zeros of the partition function for a finite system.  As the partition function, $Z$, involves a sum of positive Boltzmann factors, there can be no real positive $\beta=1/k_BT$, $T$ being the temperature and $k_B$ the Boltzmann constant, where $Z$ may vanish.  The roots are in complex conjugate pairs in the complex-$\beta$ plane.  However, as the size of the system increases, the roots may have a limit point on the real axis which then represents the transition point.  The identification follows from the fact that a zero of the partition function translates into a singularity of the free energy $F=-k_{B}T\ln{Z}$.  In addition, how the limit point is reached in the complex plane with size, the density of zeros, and the angle at which the line of zeros impinge the real axis, are related to the universality class of the transition \cite{itzykson}.  Therefore, locating the zeros in the complex plane near the real axis provides a way of characterizing a phase transition.   There are other methods of finding the transition behaviour, e.g., from the size dependence of response functions like specific heat, susceptibility, etc., with the help of data collapse assuming finite-size scaling \cite{stanley}.  Such methods have worked nicely but often pose problems for unknown systems, with several parameters to be adjusted simultaneously.   In such situations,  the method of zeros with an underlying rigorous basis has been found to be fruitful in different problems \cite{heyl,rensburg,nagata,evans,jmsmb}.

The difficulty of finding the zeros of the partition function using numerical techniques lies in the increasing level of memory requirements in the thermodynamic limit in root-finding methods.  The zeros for large systems may appear as curves, may cover some planar region, or may occur in some other complicated fashion. Although the pattern of
the zeros can be visually appealing, it is only the zeros near the positive real axis (or the imaginary axis in the case of Dynamical Quantum Phase Transition (DQPT) \cite{heyl}) that bear useful information.  Recently, a cumulant method has been proposed to locate the leading zero for a  first order transition, taking  the zipper model
of DNA as an example \cite{flindtbig}. Although the method  seems to be well behaved in case of first order transition, it calls for more detailed analysis for cases where zeros simply do not distribute uniformly, as for continuous transitions.    Our aim in this paper is to develop an iterative scheme based on the cumulant method that can
successively determine the zeros near the real axis.   

We consider primarily a model of DNA melting that can be studied exactly using real space renormalization group, namely two interacting strands of DNA on hierarchical lattices.  This particular model shows a continuous transition whose nature can be tuned by changing the dimensionality of the lattice. Effectiveness of the cumulant method from different types of data generated such as by exact methods or from Monte Carlo simulation to see to what extent they can provide us with fruitful information.  The outline of the paper is as follows.  In Sec II, we explain the method of finding zeros based on cumulants.  Sec III summarizes the model of DNA melting on hierarchical lattices.  In this section, the melting point is first determined from the closest zeros of a few lower generations (equivalently smaller system sizes).  The second closest zeros are found with some less accuracy and finally, the exponent for diverging length is calculated.  In Sec IV, this procedure is used for hetero-DNA, where the pairing energy has a binary disorder, mimicking the two possible values of base pairings. The melting point is now sample dependent and  forms a distribution. By observing how the width of the distribution varies with the system size, we conclude that there is a unique $T_{c}$ in the thermodynamic limit.  Secs III and IV deal with data obtained from exact computations.  In
contrast, in Sec V, we apply the method with moments generated from Monte Carlo simulations for a 2-dimensional Ising model. In Sec VI, the method is applied to find imaginary zeros for a one-dimensional Ising model. A summary is given in Sec VII.

\section{\label{cumulant_method} The Cumulant Method}
For a finite discrete system, the partition function is $Z= \sum_{j}  e^{-\beta E_{j}}$, where the sum is over all states $j$ with $E_{j}$ as the energy, and as already mentioned, cannot have any real, positive $\beta$ as a zero.  The zeros in the complex-$\beta$ plane, satisfying $Z(\beta)=0$, occur in complex conjugate pairs.  We discuss here a method \cite{flindtbig} of determining the leading pair closest to the real axis, and then show how the cumulants can also be used to determine the next pair of zeros.  Determination of a few zeros would allow us to compute the critical exponents associated with the transition.

\begin{figure}[h]
\includegraphics[width=0.95\linewidth]{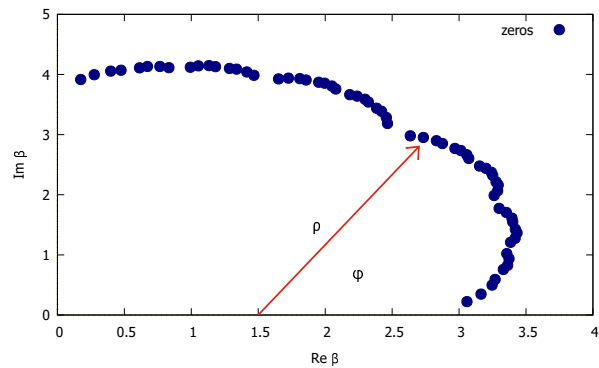}
\caption{(Color online) Schematic diagram for $\rho$ and $\phi$.The
  complex zeros are denoted by the dots.  For a chosen $\beta$ on the
  real axis, $\rho$ and $\phi$ are the polar coordinates of a zero
  from that point.}
\label{schematic1}
\end{figure}

We can define energy cumulants by taking derivatives of the logarithm of the partition function with respect to the inverse temperature $\beta$, and study the zeros in the complex-$\beta$ plane.  The $n$th order cumulant can be defined as
\begin{equation}
U_{n} = (-1)^{n} \frac{\partial^{n} \ln Z}{\partial \beta^{n}}.
\end{equation}
 According to the Weierstrass factorization theorem any entire function can be written in terms of the product of its zeros \cite{conway}. Thus, the partition function
can be written as \cite{flindtbig}
\begin{equation}
Z(\beta) = Z(0)\mathrm{e}^{\beta c}\prod_{k}\left(1-\frac{\beta}{\beta_{k}}\right),
\end{equation}
where $c$ is a constant independent of $\beta$, $\beta_{k}$ is the $k$th zero in the complex-$\beta$ plane and these zeros occur in complex conjugate pairs. Thus, the free energy becomes a sum over all the zeros
\begin{equation}
F =-\beta^{-1}\left[ \sum_{k} \ln \left(1-\frac{\beta}{\beta_{k}}\right)+\ln{Z(0)}+\beta c\right].
\end{equation}
The $n$th order cumulant is completely expressed by the zeros as
\begin{equation}
U_{n}= (-1)^{n-1}\sum_{k}\frac{(n-1)!}{(\beta_{k}-\beta)^{n}}.
\end{equation}
The cumulants are real quantities since the zeros come in complex conjugate pairs.  Eq.~$(4)$ can be further reduced to
\begin{equation}
U_{n}= 2(-1)^{n-1}(n-1)!\sum_{k}^{}\frac{\cos(n \phi_{k})}{(\rho_{k})^{n}},
\end{equation}
where $\rho_{k}$ is the distance to the $k$th zero from the point on the real $\beta$ axis at which cumulants are being calculated and $\phi_{k}$ is the angle with the real $\beta$ axis from the cumulant calculating point; see Fig.~\ref{schematic1}.  Note that the sum in Eq. $(5)$ is only over the zeros of the upper half of the complex-$\beta$ plane.  In the higher orders, cumulants will be dominated by the closest zero contribution only, i.e.,
\begin{equation}
U_{n}\approx U_{n}^{(0)}= 2(-1)^{n-1}(n-1)!\frac{\cos(n \phi_{0})}{(\rho_{0})^{n}},~n>>1.
\end{equation}

Thus, from four such successive cumulants in the higher orders we can solve for the leading pair of zeros, $\rho_{0}$ and $\phi_{0}$, using the matrix equation
\begin{equation}
\label{eq:1}
\begin{bmatrix}
    1 & -\frac{\kappa^{(+)}_{n}}{n}  \\
    1 &  -\frac{\kappa^{(+)}_{n+1}}{n+1}
 \end{bmatrix}
 \begin{bmatrix}
 -2\rho_{0}\cos\phi_{0}\\
 \rho_{0}^{2}
 \end{bmatrix}
 =\begin{bmatrix}
 (n-1)\kappa_{n}^{(-)}\\
 n\kappa_{n+1}^{(-)}
 \end{bmatrix},
\end{equation}
where, $\kappa_{n}^{(\pm)}=U^{n\pm1}/U^{n}$ \cite{flindtbig}.

Now, with the closest pair of zeros, we develop an iterative scheme to determine the next set of closest zeros which we call $(\rho_{1},\phi_{1})$.  Let us define a truncated cumulant, $\widetilde{U_{n}}$, such that $$\widetilde{U_n}=U_{n} - U_{n}^{(0)},$$ which will now be dominated by the second closest zeros in some region.  We use these $\widetilde{U_{n}}$ to define a new set of cumulant ratios $\widetilde{\kappa_{n}}$.  The procedure of Eq.~$(7)$ can then be repeated with the new set to obtain the second closest zeros $(\rho_{1},\phi_{1})$. 

Implementation of the exact method of the previous paragraph is fraught with the difficulties of propagation of errors, especially when the zeros are determined successively.  In the higher order cumulants the closest zero mostly dominates, while in the lower order other zeros should contribute too. It must be somewhere in between that the closest zero takes over and just before that the largest contribution must come from the closest and the second closest zeros.  Thus, we have to look for this crossover region, which will provide us with an upper bound on the second closest zero. The upper bound, will depend upon the error introduced in calculating the closest pair of zeros and the error propagates into the approximate cumulant calculations.  Let us consider the region where the $n$th cumulant consists mainly of the first and the second closest zeros. Then, we may write
\begin{equation}
U_{n}\approx (-1)^{n-1} (n-1)! \left[\frac{1}{(\beta_{0}-\beta)^{n}}+\frac{1}{(\beta_{1}-\beta)^{n}}\right] + c.c,
\end{equation}
where $\beta$ is the point on the real axis of the complex-$\beta$ plane at which the cumulant is being calculated, $\beta_{0}$ and $\beta_{1}$ being the closest and second closest zeros respectively. Consider,
\begin{equation}
\beta_{0}= \beta_{0}^{\prime}+\delta \beta_{0},
\end{equation}
where $\beta^{\prime}_{0}$ is the approximate value of the closest zero, and $\delta \beta_{0}$ is the error introduced when calculated from the higher order cumulants.  Then, the Eq.~$(8)$ can be written as
\begin{multline}
U_{n}\approx(-1)^{n-1} (n-1)!\Bigl[\frac{1}{(\beta^{\prime}_{0}+\delta \beta_{0}-\beta)^{n}}\\
+\frac{1}{(\beta_{1}-\beta)^{n}}\Bigr] + c.c 
\end{multline}

\begin{multline}
\approx (-1)^{n-1} (n-1)!\Bigg[\frac{1}{(\beta^{\prime}_{0}-\beta)^{n}}\left \{1-\frac{n\delta \beta_{0}}{\beta^{\prime}_{0}-\beta}\right \}\\
+ \frac{1}{(\beta_{1}-\beta)^{n}}\Bigg]+c.c.
\end{multline}
Subtracting the contribution of the closest zeros we obtain
\begin{multline}
\widetilde{U_{n}}= 2(-1)^{n-1} (n-1)!\Bigl[\frac{\cos(n\phi_{1})}{\rho_{1}^{n}}\\
-\frac{n \mid\delta \beta_{0} \mid \cos((n+1) \phi_{0}+\psi)}{\rho_{0}^{n+1}}\Bigr],
\end{multline}
where $\psi$ is the phase factor coming from $\delta \beta_{0}=\mid \delta \beta_{0} \mid \exp{(-\rm{i}\psi)}$.  Now, the first term in Eq.~$(12)$ dominates when
\begin{equation}
    n\ll\frac{\mid\ln{\mid \delta \beta_{0} \mid}\mid}{\ln{\left(\frac{\rho_{1}}{\rho_{0}}\right)}}.
\end{equation}

Eq.~$(13)$ shows how the error in the first pair of zeros puts a constraint on the order of the cumulants that can be used for the second pair.
\begin{figure}[h]
\includegraphics[width=0.5\linewidth]{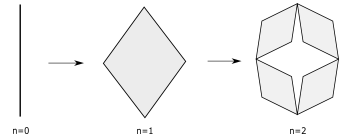}
\caption{(Color online) The recursive construction of the  hierarchical lattice for $b=2$ and $n=0,1,2...$ generations.} 
\label{lattice}
\end{figure}

\section{\label{dna_model}DNA on hierarchical lattices}
To investigate the effectiveness of the cumulant method for continuous transitions, we first consider to apply it for a model of double stranded DNA (dsDNA) on hierarchical lattice.  Hierarchical lattice with discrete scaling allows exact implementation of the real space renormalization (RG).  Thus one can write down recursion relations for the partition function and for the zeros of the partition function.  This allows us to compare the cumulant method with the exact results from RG.

On untwisting the helical structure of a dsDNA, it takes the shape of a railway track where the ties represent the bonds shared by the two strands. This structure of bond sharing can be mimicked by putting the two strands on a hierarchical lattice with the two endpoints fixed but they can wander at intermediate points.  The lattice is generated iteratively by replacing each bond in the $(n-1)$th generation with a motif of $\lambda b$ bonds to get the new $n$th generation, where $\lambda$ and $b$ are the bond scaling factor and the bond branching factor respectively; see Fig.~~\ref{lattice}.  The effective lattice dimension in the thermodynamic limit will be given by \cite{jmsmb} \begin{equation}
d=\frac{\ln \lambda b}{\ln \lambda}.
\end{equation}

In this paper we shall choose $\lambda=2$ and $b=4$ which corresponds to a three dimensional lattice.  The contact energies are defined to be $-\epsilon$ ($\epsilon > 0$) when two strands share a bond.  This constitutes our DNA model \cite{jmsmb}.

\begin{figure*}[htbp]
\begin{center}
\subfloat[]{\includegraphics[width=0.5\linewidth]{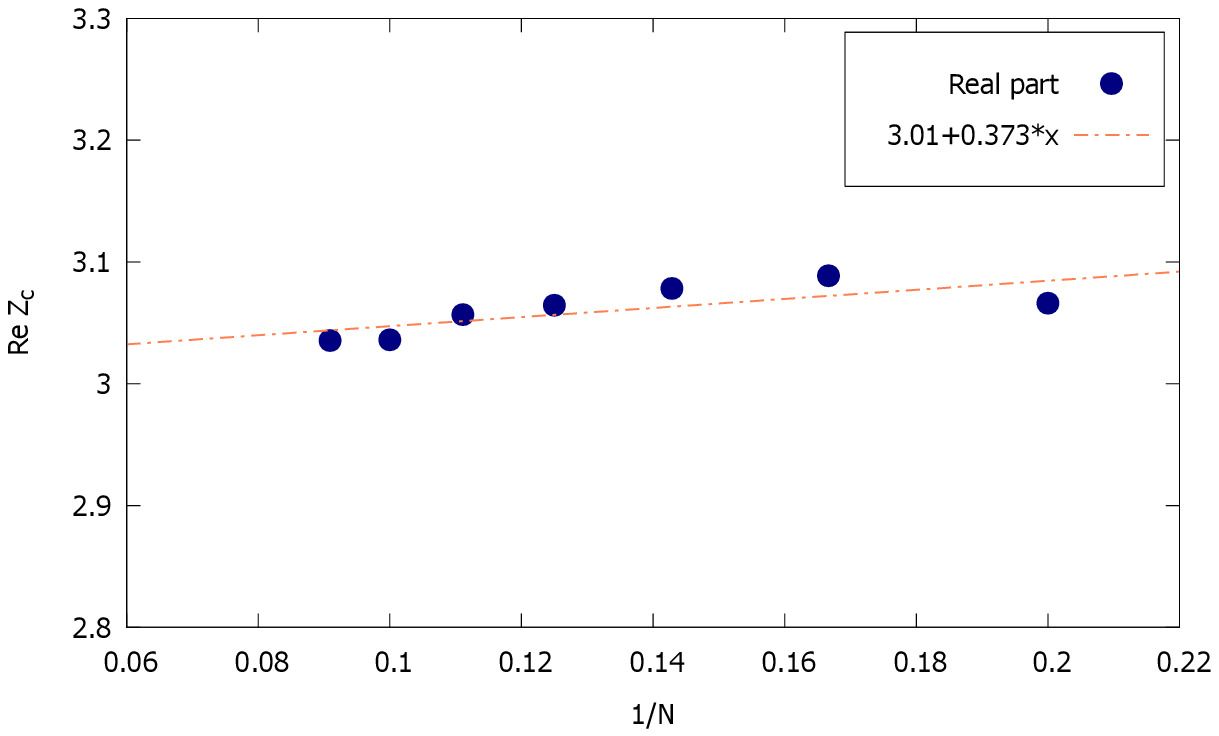}}
\subfloat[]{\includegraphics[width=0.5\linewidth]{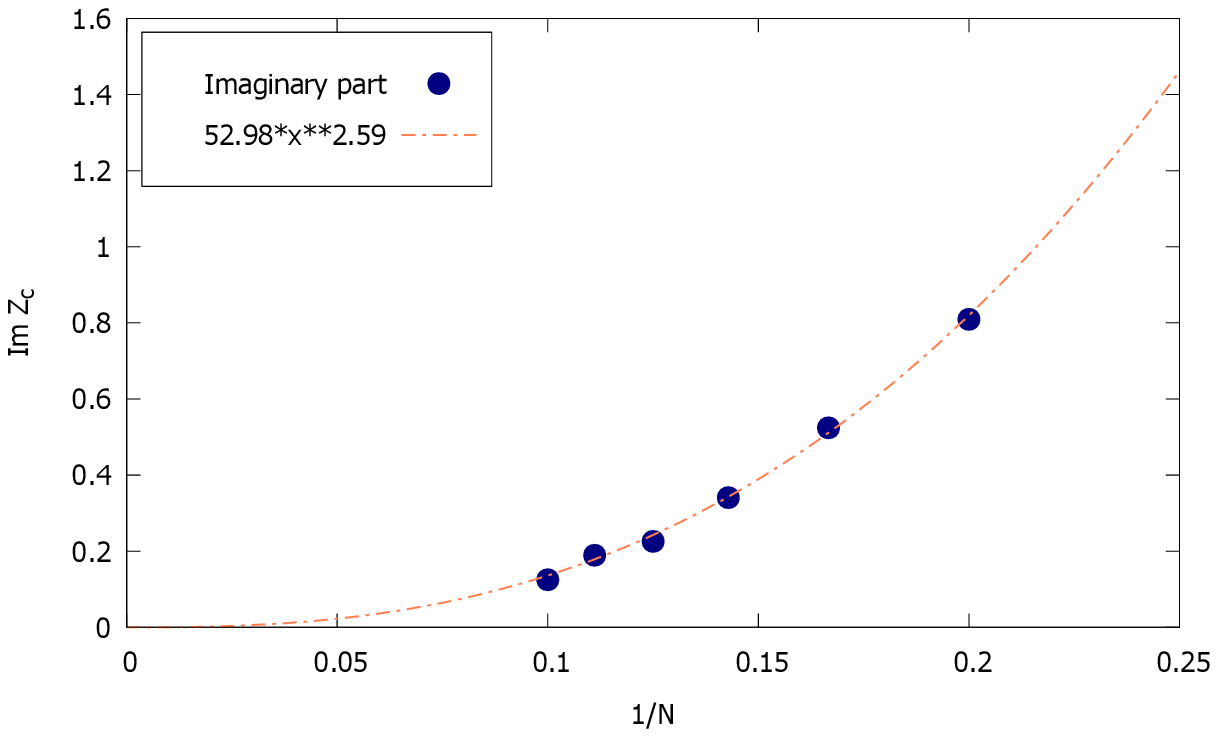}}
\end{center}
\caption{(Color online) (a) The real parts of the closest zeros for generations $N=5,6,7,8,9,10,11$.  On extrapolation to $N\to\infty$, the intersection with the $y$ axis is at $y_{c,estimate}=3.01$,
  which is close to the actual value of $y_{c,exact}=3$.  (b) The corresponding imaginary parts of the closest zeros go to zero with increase in the system size.  The cumulants were calculated at $y=2.27$.}
\label{dsDNA_closest_zeros}
\end{figure*}

\begin{figure*}[htbp]
%\centering
\includegraphics[width=0.99\linewidth]{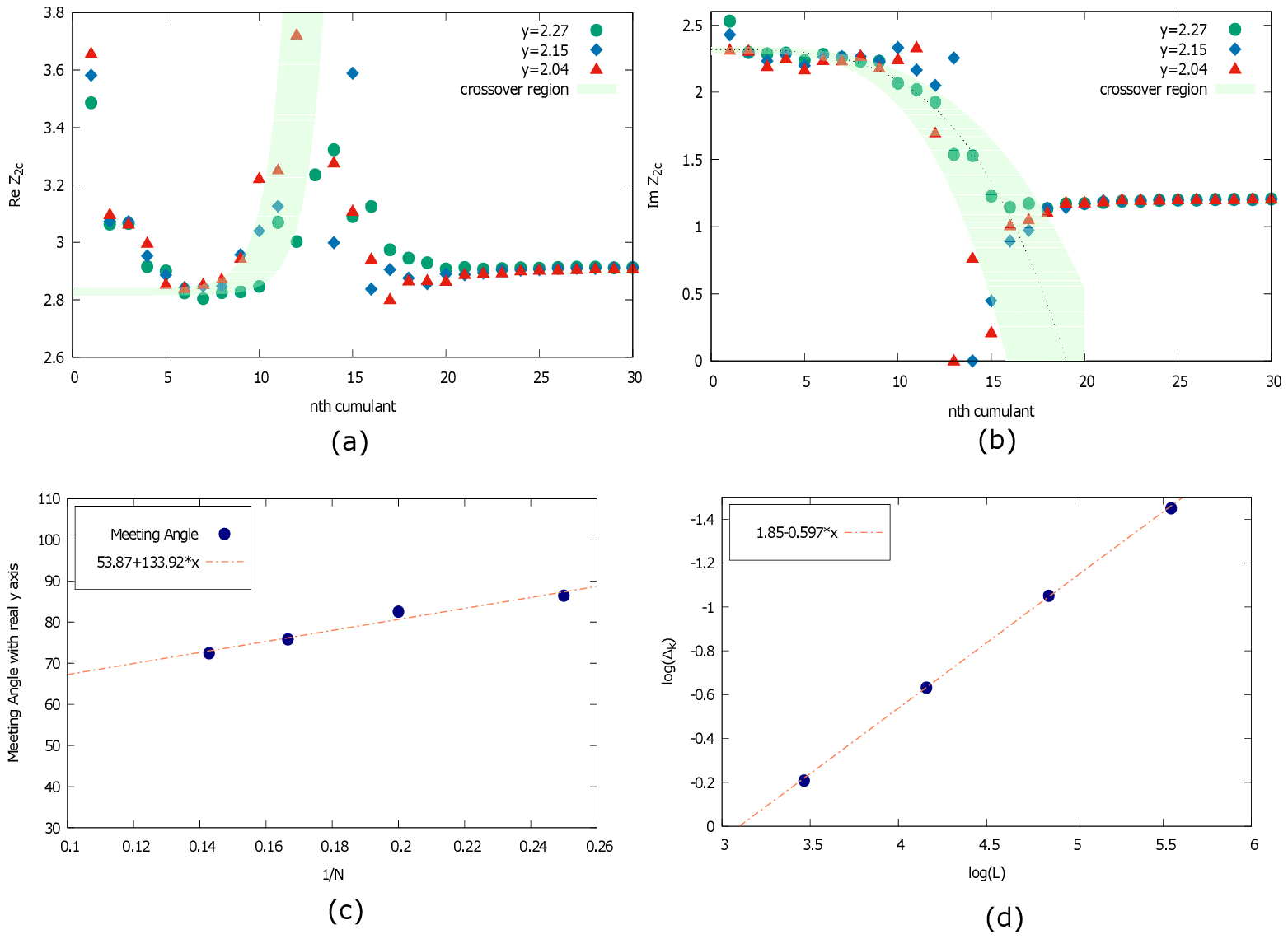}
\caption{(Color online) (a) The real part of the second closest zero of the $4$th generation from different cumulant order $n$ at three different temperatures, the outermost fitting lines encloses the
  shaded region.  (b) The imaginary part of the second closest zero of the $4$th generation from different cumulant order $n$ at three different temperatures, the outermost fitting lines encloses the
  shaded region.  (c) The angle at which the zeros meet the real $y$ axis. Estimated value is $\theta_{estimate}=53.87\degree$, the actual value is $\theta_{exact}=52.646\degree$.  (d) Estimated value
  for the critical exponent for diverging length is $\nu_{estimate}=1.67$, exact value is $\nu_{exact}=1.7095$.}
\label{dsDNA_multiplot_angles_nu}
\end{figure*}
The partition function of our model can be written in terms of the Boltzmann factor as the variable $y=\mathrm{e}^{\beta \epsilon}$, determined by the basic energy scale of the problem.   For a finite system, the partition function is a polynomial in $y$ of the form 
\begin{equation}
Z_{m} = a_{0}+a_{1}y^{2}+\cdots+a_{n}y^{2^{m}}.
\end{equation}
  We define cumulant-like quantities by taking derivatives of the logarithm of the partition function with respect to the Boltzmann factor i.e. $y$ instead of $\beta$, and study the zeros in the complex-$y$ plane.  The $n$th order cumulant can be
defined as
\begin{equation}
U_{n} = (-1)^{n} \frac{\partial^{n} \ln Z}{\partial y^{n}}.
\end{equation}
  Exact recursion relations followed by the single chain ($C_{n}$) and double chain ($Z_{n}$) partition functions can be written as  \cite{jmsmb}
\begin{equation}
C_{n}=b C_{n-1}^{2},
\end{equation}
\begin{equation}
Z_{n}=b(b-1)C_{n-1}^{4}+b Z_{n-1}^{2},
\end{equation}
with the initial conditions taken as
\begin{equation}
C_{0}=1,~~Z_{0}=y.
\end{equation}
The zeros of the partition function thus follows the following
recursion relation
\begin{equation}
q=\pm\sqrt{b\tilde{q_{j}}-(b-1)}, 
\end{equation}
where $\tilde{q_{j}}$ is one of the zero of the $(n-1)$th generation, giving rise to two new zeros in the $n$th generation \cite{jmsmb}.

It is known that dsDNA undergoes a phase transition from a bound state to an unbound state (dsDNA melting) under variation of temperature, the critical point being known as the duplex melting
point.  From RG relations the critical point is found to be $y_{c}=b-1$, with $y$ being the Boltzmann factor.

We used the method of Sec. \ref{cumulant_method} to determine the closest zeros.  To do so, the cumulants were calculated at $y=2.27$ and Eq.~$(7)$ used to determine the real and imaginary parts of the closest zeros.  The results obtained for a few smaller generations $N=5~ \text{to}~ 11$ are shown in Fig.~\ref{dsDNA_closest_zeros}.  As the figure shows, on extrapolation to infinite size, the values are in good agreement with the known exact results.  The iterative scheme of Sec. \ref{cumulant_method} is then used to determine the second closest zero.  In Fig.~\ref{dsDNA_multiplot_angles_nu}(a) and Fig.~\ref{dsDNA_multiplot_angles_nu}(b) we plot the real and imaginary parts of the second closest zero calculated at three different temperatures.  Clearly, we did not consider very low order points since all the zeros contribute there.  There appears to be a crossover region after which there is a constant contribution which corresponds
to the first closest zero. In between, there is a flat plateau which is the region of our interest. Thus, we fit our curve such as to get all the points before the crossover region and up to the flat plateau to find our best value of the second closest zero.  The thickness of the shaded region in the flat plateau should serve as an error bar to our value.  While the closest zeros can be determined with a decent accuracy of four decimal places for smaller generations, the next to closest pair of zeros can be calculated with somewhat less accuracy of one place after the decimal.

Next, we find the angle at which the zeros meet the real axis in the thermodynamic limit; see Fig.~~\ref{dsDNA_multiplot_angles_nu}(c). For this, we have taken the average of the angles made by the first and second closest zeros with the critical point (since the zeros do not fall in a smooth straight line, in lower generations) and plotted it with the generation number inverse.  Our estimate for the meeting angle of the zeros with the real $y$ axis $\theta_{estimate}=53.87\degree$, compares well with the exact result $\theta_{exact}=52.646\degree$.  The results vary substantially if we
do not include the second closest zero.

This angle contains the information about the order of the phase transition. The angle at which the zeros meet the real $y$ axis is related to the specific heat exponent as
\begin{equation}
\tan(\theta\nu)=-\tan(\pi\alpha)+\frac{A_{-}}{A_{+}}\csc(\pi\alpha),
\end{equation}
where $\theta$ is the angle between the tangent of the zeros at the limit point and the real axis of $y$, $\nu$ is the critical exponent for diverging length, $\alpha$ is the specific heat exponent and $A_{\pm}$ are the amplitudes of the specific heat on the low and high $y$ side of the transition \cite{itzykson}. For our problem with double stranded DNA, it is known that $\frac{A_{-}}{A_{+}} \rightarrow\infty$ as $A_{+}=0$. Therefore, the angle is given by
\begin{equation}
\theta=\frac{\pi}{2\nu}.
\end{equation}
Once we have $\theta$, we can find the critical exponent $\nu$. But in general we need to know $\nu$ separately. To determine $\nu$ one can make use of the RG transformation properties of the Hamiltonian to see how the partition function would change under a scaling \cite{itzykson}, and obtain the following scaling equation of the distance of the $k$th zero from the critical point
\begin{equation}
\Delta_{k}=k^{\frac{1}{d\nu}}L^{-\frac{1}{\nu}}f^{-1},
\end{equation}

\begin{figure*}[t]
%\centering
\begin{center}
\subfloat[]{\includegraphics[width=0.5\linewidth]{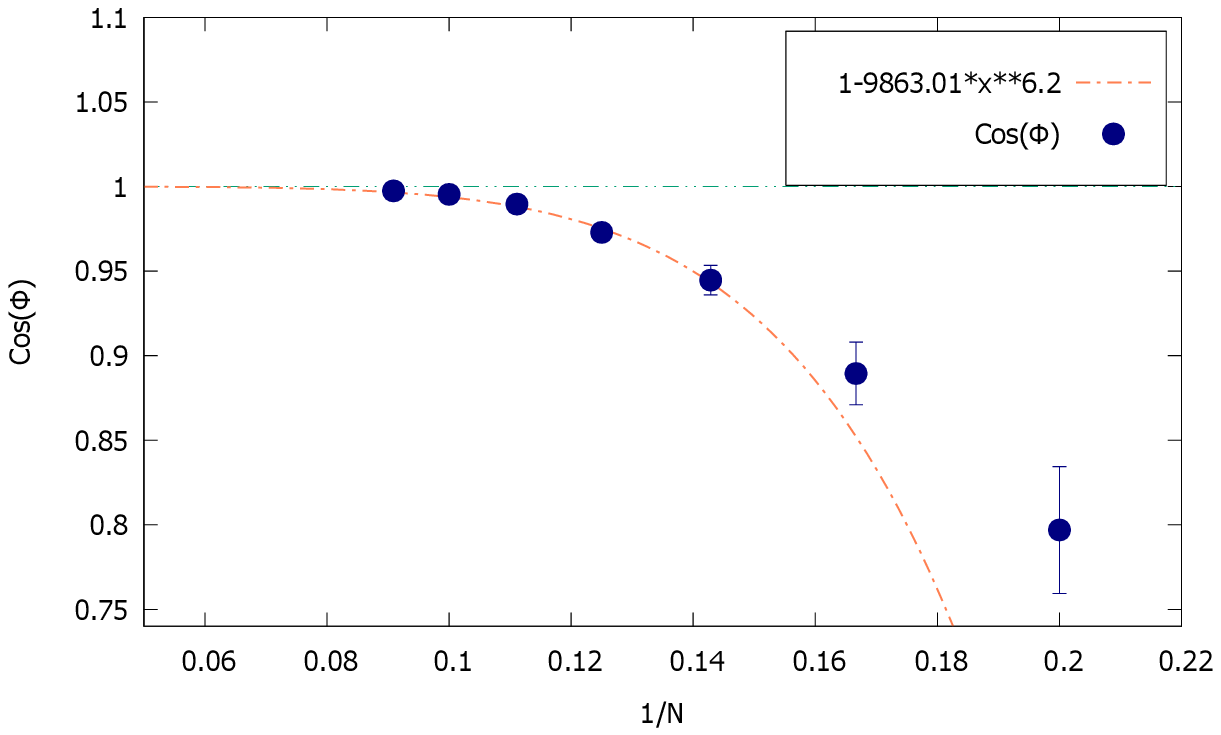}}
\subfloat[]{\includegraphics[width=0.5\linewidth]{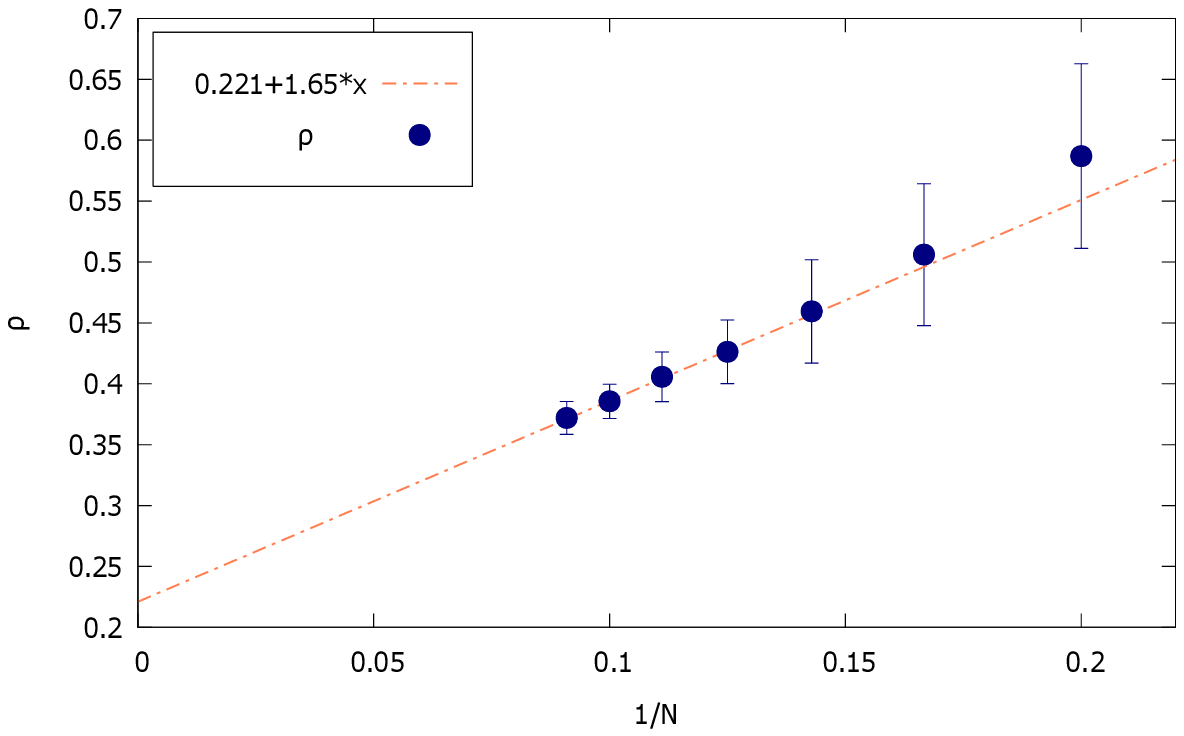}}\\
\subfloat[]{\includegraphics[width=0.33\linewidth]{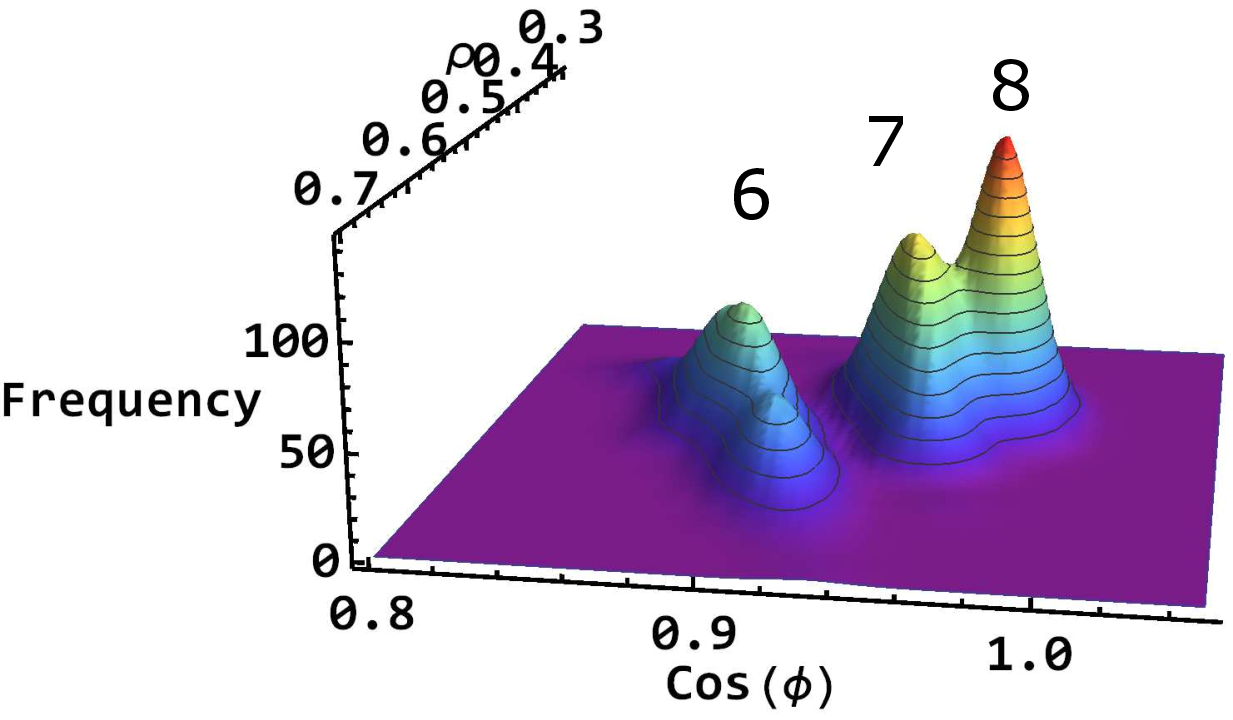}}
\subfloat[]{\includegraphics[width=0.33\linewidth]{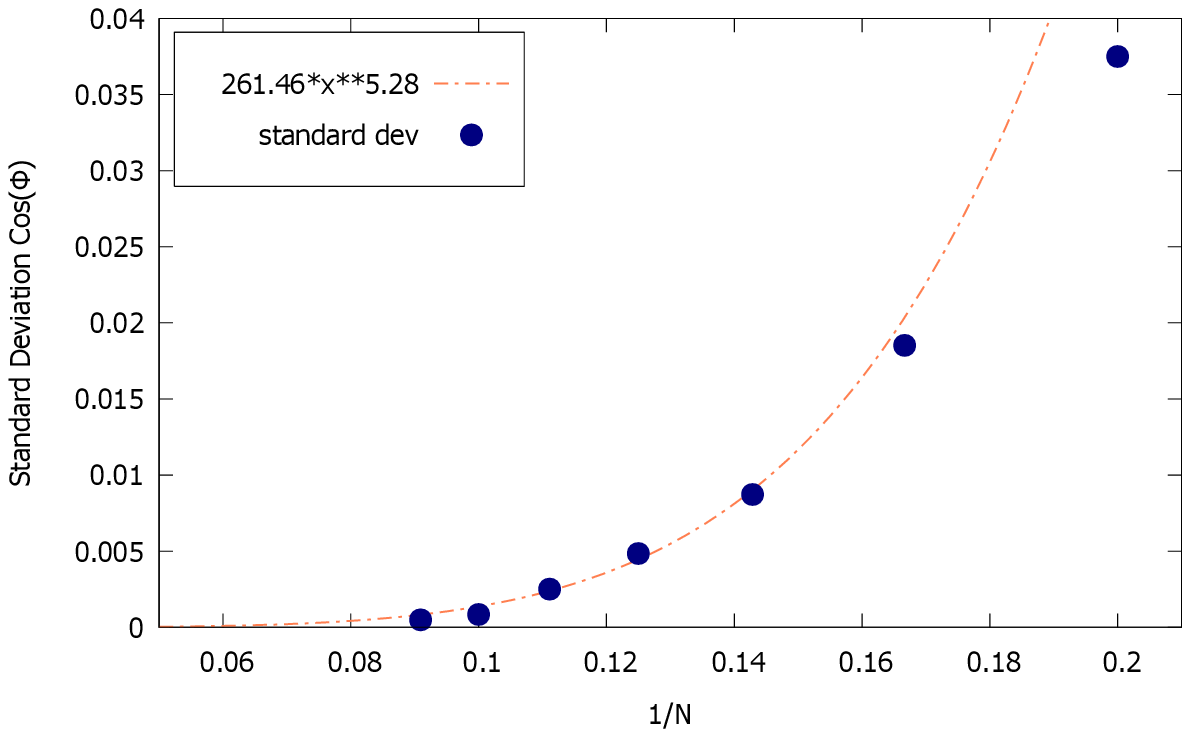}}
\subfloat[]{\includegraphics[width=0.33\linewidth]{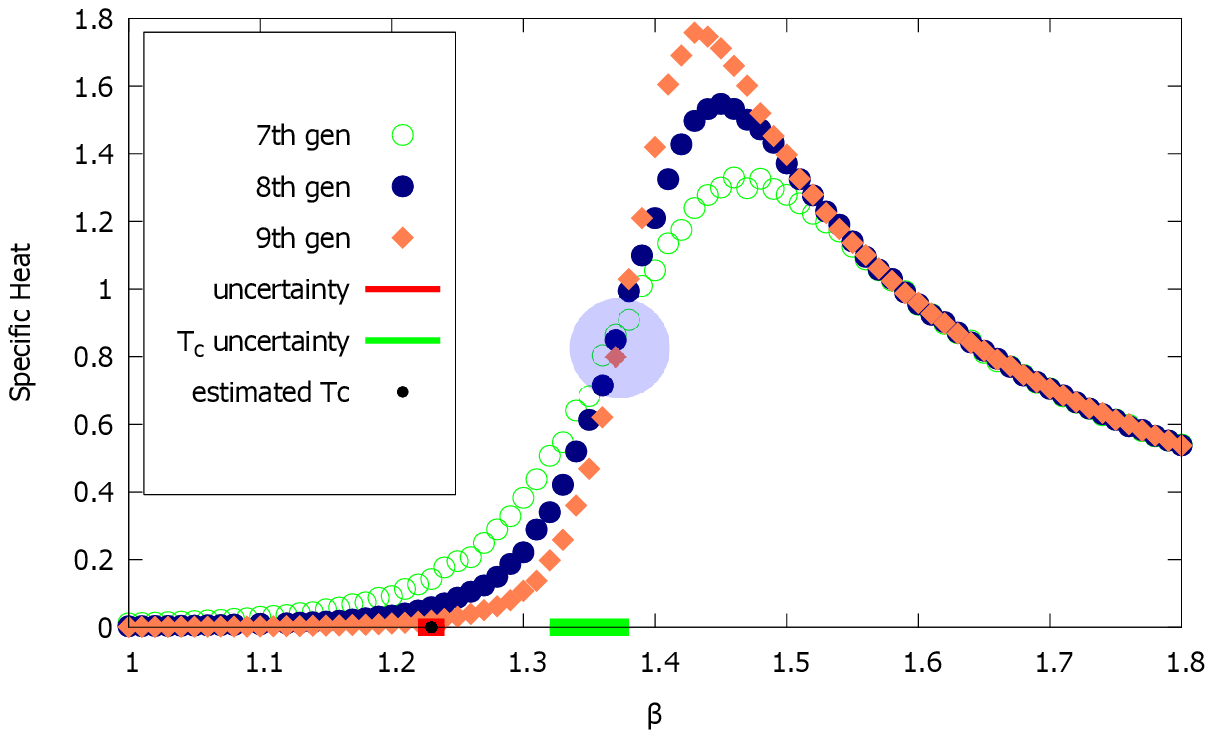}}
\end{center}
\caption{(Color online) (a) The angle made by the closest zeros from   the cumulant calculating point goes to zero in the thermodynamic limit.  The error bars for higher generations are smaller than the plotting points.  Energy cumulants are calculated at $\beta=\ln{2.75}$.  (b) Extrapolating $\rho$ using generations $N=7,8,9,10,11$,  gives the critical point at $\beta_{c}=1.23\pm0.004$. (c) $\rho$-$\cos{\phi}$ histogram for generations $N=6,7,8$. As the generation (size) increases the distribution becomes sharper. Each generation contains $200$ random samples.  (d) The fact that the width of the distribution of $\cos{\phi}$ goes to zero in the large system size limit further corroborates that there is a unique melting point. (e) The second energy cumulant $U_2$ curves scaled by the length of the corresponding generations.  Each $U_2$ curve is averaged over 500 random samples. The crossing point of the curves (transparent blue circled region) gives the quenched averaged $\beta_{c}$ indicated by  a thick green line on the $\beta$-axis. The range from the zeros is represented by a black dot and red thick line.}
\label{dsDNA_random_disorder}
\end{figure*}
where $d$ is the dimension, $\Delta_{k}$ is the distance of the $k$th zero from the critical point, $L$ is the size of the system and $f^{-1}$ is a constant. To determine $\nu$ we plot $\ln\Delta_{k}$ vs $\ln L$, for the closest zeros ($k=1$) for generations $N=5,6,7,8$; see Fig.~~\ref{dsDNA_multiplot_angles_nu}(d).

\begin{figure*}[t]
%\centering
\subfloat[]{\includegraphics[width=0.33\linewidth]{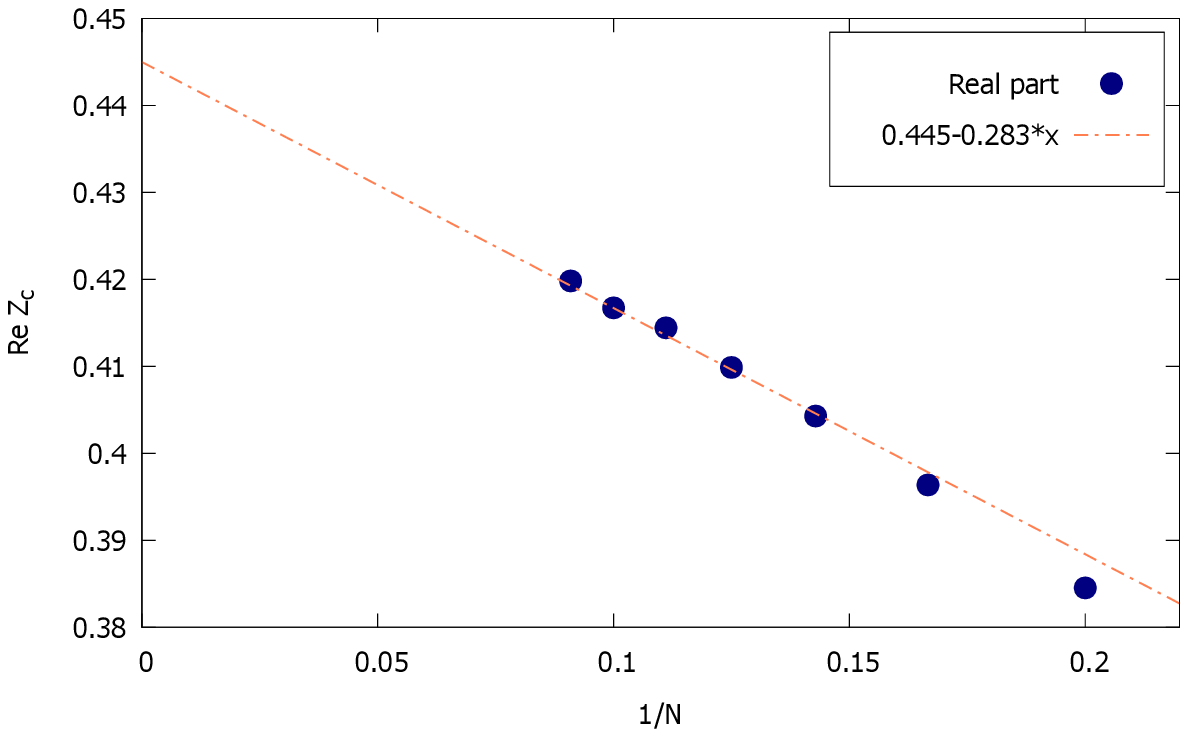}}
\subfloat[]{\includegraphics[width=0.33\linewidth]{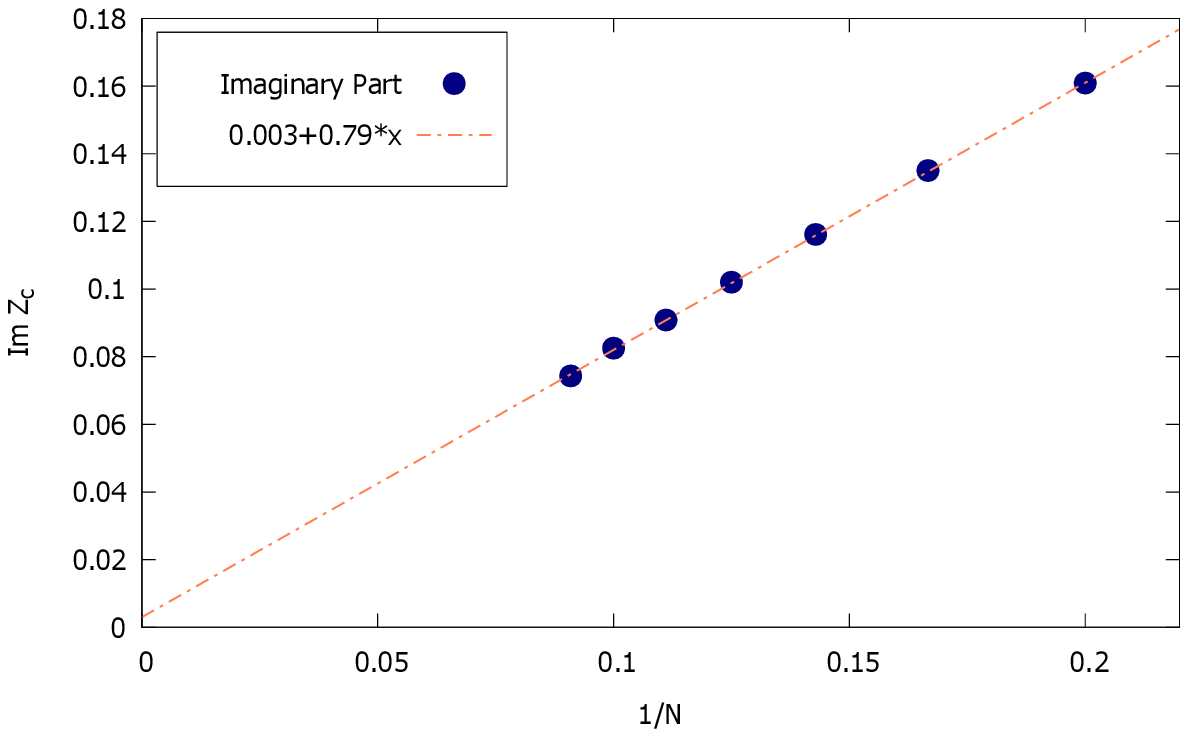}}
\subfloat[]{\includegraphics[width=0.33\linewidth]{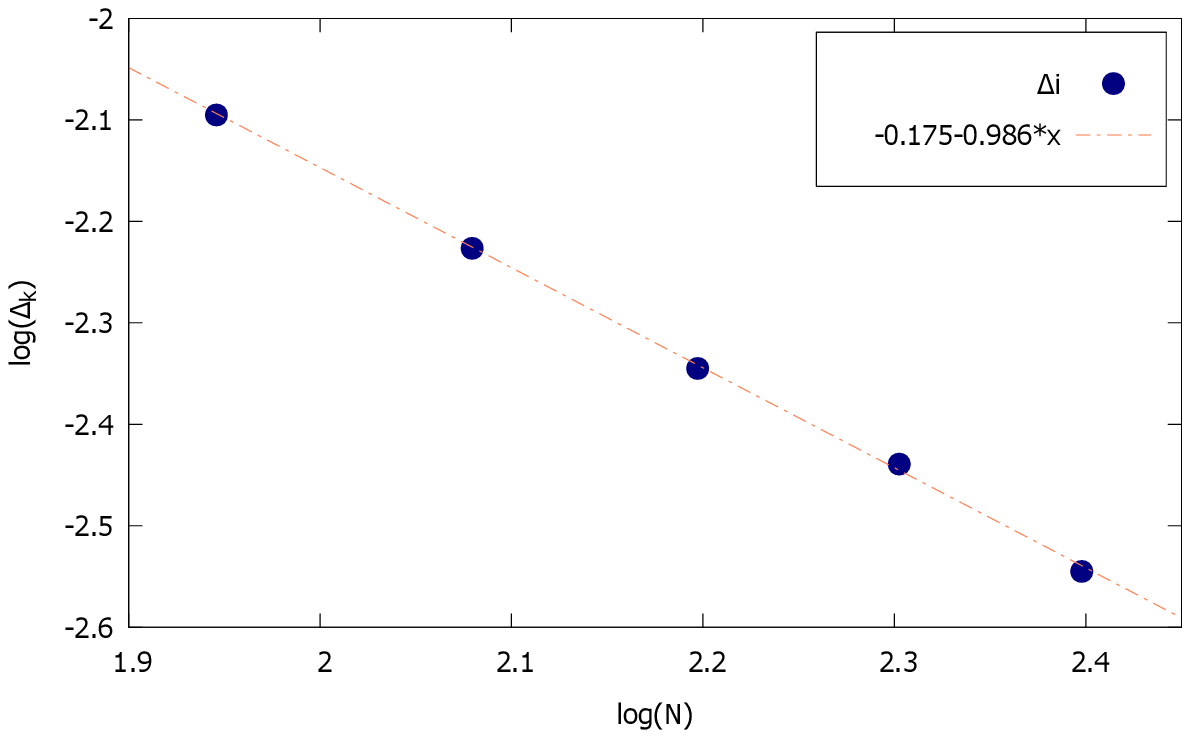}}
\caption{(Color online) (a) The meeting point of the zeros with the real $\beta$ axis on the complex-$\beta$ plane, is found from the leading order zeros of $N\times N$ square lattice with
  $N=5,6,7,8,9,10,11$, which meets at $0.445$ which is quite close to the actual value of the inverse of the critical temperature $\beta_{c}=0.44068$. (b) The imaginary part of the closest zero for
  $N\times N$ square lattice with $N=5,6,7,8,9,10,11$, approaches $0$ in the thermodynamic limit. It meets at $0.003014$. (c) The critical exponent $\nu$ for correlation length from closest zeros of $N\times N$ square lattice with $N=7,8,9,10,11$, found out to be $\nu_{estimated}=1.01$, the exact value is $\nu_{exact}=1$.  The cumulants are calculated at $\beta=1/3$ with $J/k_{B}=1$.}
\label{2d_ising}
\end{figure*}
Our estimate of the length exponent $\nu_{estimate}=1.67$ compares reasonably well with the independent estimate from  Eq.~$(22)$ which also gives $\nu=1.67$.  The exact value of $\nu_{exact}=1.70951$.  It might be possible to improve the accuracy of the result by iterative inclusion of other zeros.

\section{\label{random_bonds}Effects of  binary random disorder}
In order to show the usefulness of the cumulant method, we now use it for a heterogeneous DNA.  The non-uniformity in the base sequence of a DNA can be handled by specifying the base pair energies along the chain. A simpler situation is a random sequence where the interaction energy is randomly chosen from a specified distribution \cite{smsmb1,smsmb2,smsmb3,monthus}.

With the introduction of randomness, the obvious question is whether a critical point for melting exists. Since the partition function is different for different realization of randomness, any thermally averaged quantity will have a probability distribution.  One such important distribution is the probability distribution of the closest zero. In the thermodynamic limit, a well-defined
melting point would require a sharp probability distribution.  The width of the distribution should vanish in the large length limit ("self-averaging").  Although, that is not enough for a transition to occur the zeros must touch the real axis.  However, this is not equivalent to the traditional quenched averaging.  The quenched average transition temperature can be determined from the average
specific heat as we discussed below.

As a model system, we take two different types of pairing energy, $\epsilon_{1}=1.0$ and $\epsilon_{2}=0.5$  with equal probability.  The randomness depends only on the longitudinal direction and has no dependence on the transverse direction.  For the closest zeros of each sample,  $\rho$ and $\cos\phi$ are then determined by the using the method of Sec. \ref{cumulant_method}  from the four successive cumulants of order $n=19,20,21,22$.  The cumulants were calculated at $\beta=\ln{2.75}$ and the whole procedure is repeated for $200$ random samples.  Fig.~~\ref{dsDNA_random_disorder}(a) shows that the average  $\overline{\cos{\phi}}$, where the overbar represents an averaging  over samples, approaches $1$ on extrapolation.  This  indicates that there is a limit point on the real $\beta$ axis. see A similar extrapolation of $\overline{\rho}$ in Fig.~~\ref{dsDNA_random_disorder}(b) locates the transition point at  $\beta_{c}=1.23\pm0.004$.

Fig.~~\ref{dsDNA_random_disorder}(c) and Fig.~~\ref{dsDNA_random_disorder}(d) show the distributions of $\rho$ and $\cos{\phi}$ for successive generations of $N=6,7,8$.  The width of the joint distribution is found to go to zero in the thermodynamic limit.  An independent way of determining the transition temperature is to consider the average specific heat or the second energy cumulant $U_2$ for various sizes of the system and look for the common crossing point.  Fig.~~\ref{dsDNA_random_disorder}(e) shows the second energy cumulant $U_2$ curves (scaled by the length of DNA) for three different generations, $N=7,8,9$, each one averaged over $500$ samples. The intersection of the $U_2$ curves gives the melting temperature to be $\beta_{c}=1.35\pm0.03$ (transparent blue circled region), where the uncertainty is based on the sample fluctuations of the intersection point.  For comparison, the melting point for a pure system with $\epsilon_1$ is $\beta_c(\epsilon=1)=1.0986$, while for the other one, $\beta_c(\epsilon=0.5)=2.1972$. The zeros seem to be dominated by the tighter pairs.  For small lengths, it is possible to get samples with over representation of the lighter energy bonds that led to a double peaked distribution.  As there are situations like Griffiths' phase \cite{griffiths} where zeros of rare regions play a crucial role, this cumulant method of finding zeros will be relevant there.

\section{\label{ising2d}Ising 2d}
In the previous sections \ref{dna_model} and \ref{random_bonds}, we determined the closest zeros as well as the second closest zeros from the higher energy like cumulants of the partition function. There we had the privilege of finding the exact cumulants since the partition function is known exactly. But in simulations, the higher energy cumulants can be calculated more easily, albeit approximately, than finding the partition function.  One such example is the two dimensional Ising model. The Hamiltonian for 2d Ising model with interaction between the nearest neighbours in the absence of an external magnetic field is
\begin{equation}
    \beta\mathcal{H}=-J\sum_{\langle ij\rangle}\sigma_{i}\sigma_{j},
\end{equation}
where $\sigma_{i}\in \{-1,1\}$ is the spin at site $i$, $J$ is the coupling constant between neighboring spins and $\langle ij\rangle$ indicates nearest neighbors.  It is known that for such a two-dimensional spin system there is a phase transition from paramagnetic to ferromagnetic phase at a finite temperature.  The energy moments were obtained from Monte-Carlo simulation using the Metropolis algorithm \cite{montecarlo}  and then  {\small MATHEMATICA} was used to determine the cumulant \cite{mathematica}. Since our aim throughout this work had been to determine quantities in the thermodynamic limit by studying the zeros of small system size, we have calculated the leading zeros for $N\times N$ square lattices with $N = 5,6,7,8,9,10,11$ using the cumulant method.  To find the zeros we have calculated up to $30$th moment of energy using quadruple precision data type at $\beta=1/3$ with $J/k_{B}=1$.

\begin{figure*}[htbp]
%\centering
\includegraphics[width=0.99\linewidth]{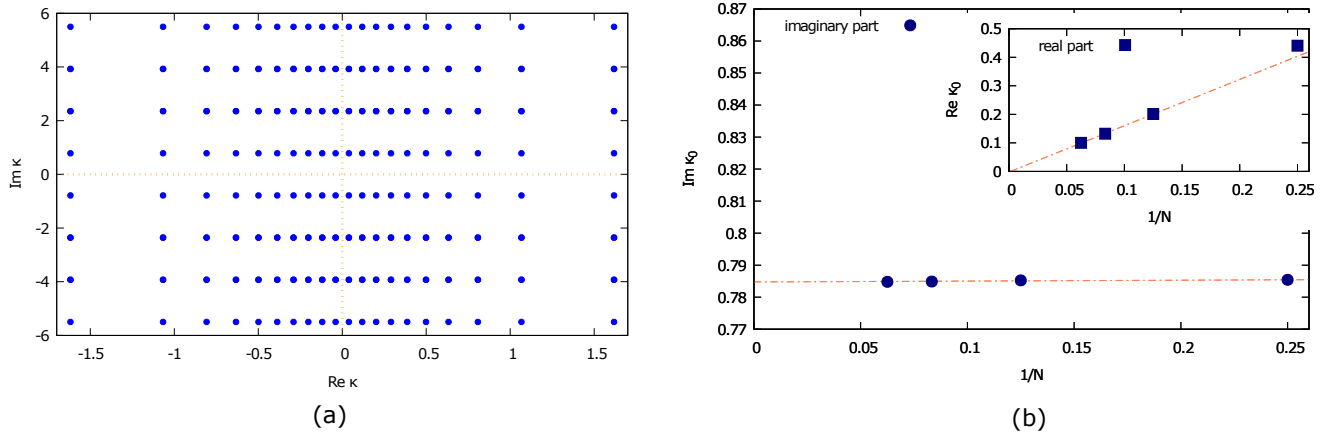}
\captionof{figure}{(Color online) (a) Phase transition happens periodically along the imaginary temperature axis.  Periodical transitions along imaginary temperature axis shown for $N=40$.  (b) Zeros of the 1-d Ising model in the complex-$\kappa$ plane for $N=4,8,12,16$. Cumulants calculated at $\kappa=0.5\rm{i}$.}
\label{1d_ising}
\end{figure*}

\begin{center}
\begin{tabular}{ |l|c|c|c| }
\hline
 & $\Re \beta_{c}$ & $\Im \beta_{c}$ & $ \nu $ \\ \hline
\multirow{3}{*}{Estimated}
 & & &\\
 & $ 0.445 $ & $ 0.003 $ & $ 1.01 $ \\
 & & &\\

 \hline
\multirow{3}{*}{Exact}
 & & &\\
 & $ 0.44068 $ & $0$ & $ 1 $ \\
 & & &\\
 \hline
 \end{tabular}
\captionof{table}{Comparison with exact values for 2d Ising model.}
\label{ising_table}
\end{center}

Fig.~~\ref{2d_ising}(a,b) show the real and the imaginary parts of the leading zero.  Fig.~~\ref{2d_ising}(b) shows that the imaginary part of the closest zero goes to zero in the thermodynamic limit as required for a phase transition.  The critical temperature is then given by the extrapolated value in Fig.~~\ref{2d_ising}(a). Fig.~~\ref{2d_ising}(c) shows how the distance between the closest
zeros and the transition point scales with the system size.  The inverse of the slope of the fitting line gives the critical exponent $\nu$ for the diverging correlation length according to the Eq~$(23)$.  In Table \ref{ising_table} we have compared our numerical results with the exact results in the thermodynamic limit.

\section{\label{ising1d}Imaginary zeros in Ising 1d}

Motivated by the fact that a phase transition can happen with time as a parameter in case of DQPT where the zeros meet on the imaginary axis \cite{heyl}, we have found the zeros closest to the imaginary axis, for the 1d Ising model which do not have any transition at any non zero temperature.  The partition function of the 1d Ising model is exactly known to be
\begin{equation}
    Z = \lambda_{+}^{N}+\lambda_{-}^{N},
\end{equation}
where $\lambda_{\pm}$ are the eigenvalues of the transfer matrix and $N$ is the number of spins.

In absence of any external magnetic field, the eigenvalues become
\begin{equation}
    \lambda_{\pm}=e^{\kappa}\pm e^{-\kappa},
\end{equation}
where $\kappa=\frac{J}{k_{B}T}$ and the zeros of the partition function on the complex-$\kappa$ plane are found to be
\begin{equation}
\label{eq:2}
    \kappa_j=\frac{1}{2}\log\left[-\rm{i}\tan{\left(\frac{(2j+1)\pi}{2N}\right)}\right],
\end{equation}
with $j=0,1,2\cdots N-1$. We here consider only even values of $N$ for which the zeros in the complex-$\kappa$ plane are distributed symmetrically about the imaginary axis.   The zeros meet the
imaginary axis periodically due to the presence of the $-\rm{i}$ factor inside log in Eq.~$(27)$; see Fig.~~\ref{1d_ising}(a).

Assuming a symmetric distribution about the imaginary axis, we can define a free energy like quantity as 
\begin{equation}
    F=\ln{(Z Z^{*})},
\end{equation}
where $Z$ and $Z^{*}$ are the partition functions evaluated at a complex temperature and its complex conjugate respectively so that $F$ is real.  For pure imaginary $\kappa$, there is a first order transition at $|\kappa|=\pi/4$, where $|\lambda_+|=|\lambda_{-}|$, and odd multiples of it.  These transition points are the zeros given by Eq. (\ref{eq:2}), and are related to
DQPT \cite{heyl}.

In terms of zeros of the complex-$\kappa$ plane
\begin{equation*}
    F=\ln{\left[\prod_{j}(\kappa_{j}-i\kappa) \prod_{l}(\kappa_{l}+i\kappa)\right]}.
\end{equation*}
The $n$-th cumulant is given by
\begin{eqnarray}
 U_{n}&=&(-\partial_{i\kappa})^{n}\ln (ZZ^{*})\\
 &=&(-\partial_{i\kappa})^{n}\ln{\left[\prod_{j}(\kappa_{j}-i\kappa) \prod_{l}(\kappa_{l}+i\kappa)\right]}.
 \end{eqnarray}
Retaining only the terms of the closest zeros, the above equation can be written as
\begin{eqnarray}
 U_{n}=4(-1)^{n-1}(n-1)!\frac{\cos{(n\phi_{0})}}{(i\rho_{0})^{n}},
\end{eqnarray}
which requires a little modification of the matrix equation Eq. \ref{eq:1} by simply putting $i\rho_{0}$ instead of $\rho_{0}$. Here, $\phi_{0}$ is the angle made with the imaginary axis by the vector
extending from the cumulant calculating point on the imaginary axis.  Because of the periodic pattern of the zeros, care is needed in choosing the cumulant calculating point.  Fig.~~\ref{1d_ising}(b) shows the real and imaginary parts of the closest zeros of the branch closest to the real $\kappa$ axis on the upper half of the complex-$\kappa$ plane, for different system sizes.  We do recover the closest zero properly.

\section{\label{sec:level5}Summary}
To summarize, we developed an iterative scheme to determine the zeros of a partition function using cumulant data. The method has been tested for DNA melting transition on a hierarchical lattice for  which the zeros can be determined  exactly.  We  found the leading pair of zeros of the partition function both for pure and hetero-DNA cases.  The next leading pair of zeros are also found which proved to be important in determining the angle at which the zeros approach the real axis and also in determining the critical exponent.  We have chosen such a system to
have a good comparison with the exact results. The results found out to be matching well enough with the theoretical exact results, thus giving us a new way to look into systems in the thermodynamic limit from smaller system sizes. Although the zeros from the lower generations may not always be helpful, still it proved to be quite a powerful method. Moreover, it is shown that the method is well-suited for results from Monte Carlo simulation proved to be equally good to reproduce the known results for the critical exponent and critical point estimation for the case of Ising model in 2d.  An extension of the method has been shown to be useful for the imaginary zeros too.

\section*{\label{sec:level51}Acknowledgements} 
D.M. would like to thank Jaya Maji and Goutam Tripathy for useful discussions.   Computations performed on the SAMKHYA high performance computing facility at Institute Of Physics Bhubaneswar.

\end{document}